\documentclass[apj]{emulateapj}
\usepackage{natbib}
\usepackage{apjfonts}
\usepackage{graphicx}

\newcommand{\xmm}{{\em XMM-Newton}}
\newcommand{\chandra}{{\em Chandra}}
\newcommand{\spitzer}{{\em Spitzer}}
\newcommand{\hst}{{\em Hubble}}
\newcommand{\vla}{{\em VLA}}
\newcommand{\rosat}{{\em ROSAT}}

\newcommand{\sersic}{S\'{e}rsic~}
\newcommand{\ergsscm}{~ergs~s$^{-1}$~cm$^{-2}$}

\newcommand{\ergss}{~ergs~s$^{-1}$}
\newcommand{\kms}{~km~s$^{-1}$}

\newcommand{\um}{~$\mu$m}
\newcommand{\msun}{{M$_{\odot}$}}
\newcommand{\lsun}{{L$_{\odot}$}}
\newcommand{\cs}{{c$_{s}$}}
\newcommand{\dg}{$^{\circ}$}

\bibpunct{(}{)}{,}{a}{}{}

\begin{document}
\slugcomment{Accepted for publication in The Astrophysical Journal}

\title{Constraining the Outburst Properties of the SMBH in Fornax A through X-ray,
Infrared, and Radio Observations}

\shorttitle{Fornax A: Multiwavelength Constraints} 
\shortauthors{Lanz et al.}
\author{Lauranne~Lanz\altaffilmark{1}}
\author{Christine~Jones\altaffilmark{1}}
\author{William~R.~Forman\altaffilmark{1}}
\author{Matthew~L.~N. ~Ashby\altaffilmark{1}}
\author{Ralph~Kraft\altaffilmark{1}}
\author{Ryan~C.~Hickox\altaffilmark{1,2}}

\altaffiltext{1}{Harvard-Smithsonian Center for Astrophysics, 60 Garden St., Cambridge,
MA 02138, USA; llanz@head.cfa.harvard.edu}
\altaffiltext{2}{Department of Physics, Durham University, Durham DH1 3LE, UK}

\begin{abstract}

Combined \spitzer, \chandra, \xmm, and \vla~observations of the
giant radio galaxy NGC 1316 (Fornax A) show a radio jet and X-ray 
cavities from AGN outbursts most likely triggered by a merger 
with a late-type galaxy at least 0.4~Gyr ago. We detect a weak 
nucleus with an SED typical of a low-luminosity AGN with a 
bolometric luminosity of $2.4\times10^{42}$~\ergss. We examine the 
\spitzer~IRAC and MIPS images of NGC 1316. We find that the dust 
emission is strongest in regions with little or no radio emission 
and that the particularly large infrared luminosity relative 
to the galaxy's K-band luminosity implies an external origin 
for the dust. The inferred dust mass implies that the merger 
spiral galaxy had a stellar mass of $1-6\times10^{10}$~\msun~and 
a gas mass of $2-4\times10^{9}$~\msun. X-ray cavities in the 
\chandra~and \xmm~images likely result from the expansion of 
relativistic plasma ejected by the AGN. The soft (0.5-2.0~keV) 
\chandra~images show a small $\sim15\arcsec$ (1.6~kpc) cavity 
coincident with the radio jet, while the \xmm~image shows two 
large X-ray cavities lying 320$\arcsec$ (34.8~kpc) east and 
west of the nucleus, each approximately 230$\arcsec$ 
(25 kpc) in radius. Current radio observations do not show
emission within these cavities. The radio lobes lie at radii 
of $14\farcm3$ (93.3 kpc) and $15\farcm6$ (101 kpc), more distant 
from the nucleus than the detected X-ray cavities. The relative morphology 
of the large scale 1.4~GHz and X-ray emission suggests they were 
products of two distinct outbursts, an earlier one creating the 
radio lobes and a later one producing the X-ray cavities. 
Alternatively, if a single outburst created both the X-ray 
cavities and the radio lobes, this would require that the radio
morphology is not fully defined by the 1.4~GHz emission. For 
the more likely two outburst scenario, we use the buoyancy rise times 
to estimate an age for the more recent outburst that created them 
of 0.1~Gyr and the $PV$ work done by the expanding plasma to
create the X-ray cavities to estimate the outburst's energy to be 
$10^{58}$~ergs. The present size and location of the radio lobes 
implies that the outburst that created them happened $\sim0.4$~Gyr 
ago and released $\sim5\times10^{58}$~ergs.

\end{abstract}


\section{Introduction}

NGC 1316 (Fornax A) is one of the nearest and
brightest radio galaxies with radio lobes 
spanning 33$\arcmin$ \citep{eke83}. It lies on the outskirts of
the Fornax cluster at a distance of 22.7$\pm$1.8~Mpc, based 
on a distance modulus of 31.66$\pm$0.17 
\citep{ton01}.\footnote{We adopt $h=0.70$ and therefore revise
the \citet{ton01} distance which used $h=0.74$. This gives a scale of 
$9\farcs2$~kpc$^{-1}$ at the distance of NGC 1316.} Early 
optical observations of NGC 1316 \citep{eva49} revealed
dust lanes in the nuclear region, leading \citet{shk62}
to hypothesize that radio lobes might 
be powered by accretion of interstellar gas onto the 
nucleus. More extensive optical observations led \citet{sch80} 
to classify NGC 1316 as a D-type galaxy \citep{mor58}
with an elliptical-like spheroid embedded in a large envelope.

Schweizer further suggested that NGC 1316's disturbed morphology 
may be due to one or more low-mass gas-rich mergers, 
occurring over the last 2~Gyr. \citet{mac98} concluded, 
from more recent optical observations, in 
combination with~\rosat~data, that NGC 1316 had undergone either
a major merger more than 1~Gyr 
ago or a merger with a low-mass gas-rich galaxy about
$\sim$~0.5~Gyr ago. NGC 1316 exhibits other signs of
at least one merger, including loops of H$\alpha$ filaments 
resembling tidal tails with projected lengths as large as 
10$\arcmin$ (65.4~kpc) and a gas disk rotating much faster than the stellar
spheroid and at an angle to it, indicating a likely external origin
\citep{xil04, sch80}. \citet{gou01} used
globular clusters to date any major mergers to between 1.5 and 4~Gyr ago, but
the age and type of merger (or mergers) still remain uncertain.

Multiwavelength observations can set constraints
on the merger event. Radio emission, from the
radio jet and lobes, is expected to be powered 
by accretion onto the central supermassive black hole 
(SMBH), which can be enhanced by the infall of material 
from a merger.  The expanding radio plasma can create 
cavities in the surrounding hot gas, which are seen as 
decrements in the X-ray emission. By measuring
the $PV$ work done by the expanding radio plasma \citep{mcn00,chu02}, 
we can constrain the energy produced by the SMBH. Also, since the 
X-ray cavities rise buoyantly, their distance from the nucleus constrains 
the age of the outburst. We can set mass-related constraints on 
the merger galaxy by examining NGC 1316 for dust. As an early-type 
galaxy, NGC 1316 is expected to be dust poor. Mid-infrared 
observations permit us to measure the amount of dust present, 
which, if higher than expected for a galaxy of its size and type, 
indicates an external origin for the dust. 

In this paper, we report our analysis of the surface 
brightness distribution of NGC 1316 in the mid-IR with the 
Infrared Array Camera (IRAC) and the Multiband Imaging
Photometer for Spitzer (MIPS) of the \spitzer~$Space~Telescope$ and the 
resulting determination of the warm dust morphology. We describe our
analysis of the X-ray emission imaged both by \chandra~and \xmm. Observations
and data reduction are described in \S 2, and imaging and modeling results are
given in \S 3. We compare the features seen in the infrared and X-ray with the
radio emission in the nuclear, inner jet, and extended emission regions in
\S 4 and examine the constraints these observations 
place on the mass of the merger progenitor, the outburst and merger ages, and 
the outburst energies in \S 5. Finally, we summarize the 
results in \S 6. Images have north to the top and east to the left. Angles are
given counter-clockwise from west, unless otherwise stated. 

\section{Observations and Data Reduction}

\subsection{\spitzer~Observations}
IRAC \citep{faz04} observations of NGC 1316 were 
obtained on 2004 July 19 and 22 as part of the SIRTF Nearby
Galaxy Survey (SINGS) Legacy 
program (Kennicutt et al. 2003; \spitzer~PID 159) in all four 
bands. The two visits to NGC 1316 consisted of similar $4 \times 30$~s 
integrations covering the galaxy, its companion
NGC 1317, and the nearby field. 

For our analysis, we retrieved the Basic 
Calibrated Data (BCD) version S14.0 pipeline products from the 
\spitzer~archive. We corrected the IRAC BCD frames 
for residual images arising from prior, unrelated 
observations of bright sources by making object-masked, 
median-stacked coadds of all science frames not 
containing significant extended emission from NGC 1316 
or NGC 1317. These single-visit coadds were then 
subtracted from the individual BCDs of the 
corresponding visit to remove the 
residual images. For the 8.0\um~observation, 
light scattered from the galaxy nucleus along the 
detector array rows was fit and subtracted from the 
BCD frames using custom software as in \citet{ash09}. We then coadded the 
modified 30~s BCD frames from both observations using version 4.1.2
of {\tt IRACproc} \citep{sch06} to mosaics having 
1\farcs2 pixels, i.e., the native IRAC pixel size. 

The SINGS program also obtained MIPS observations of NGC 1316
\citep{rie04} on 2004 December 5 and 7. We 
obtained a combined 24\um~mosaic and coverage map of NGC 1316 
from the SINGS 
website\footnote{http://data.spitzer.caltech.edu/popular/sings}.

\begin{figure*}
\centerline{\includegraphics[width=\linewidth]{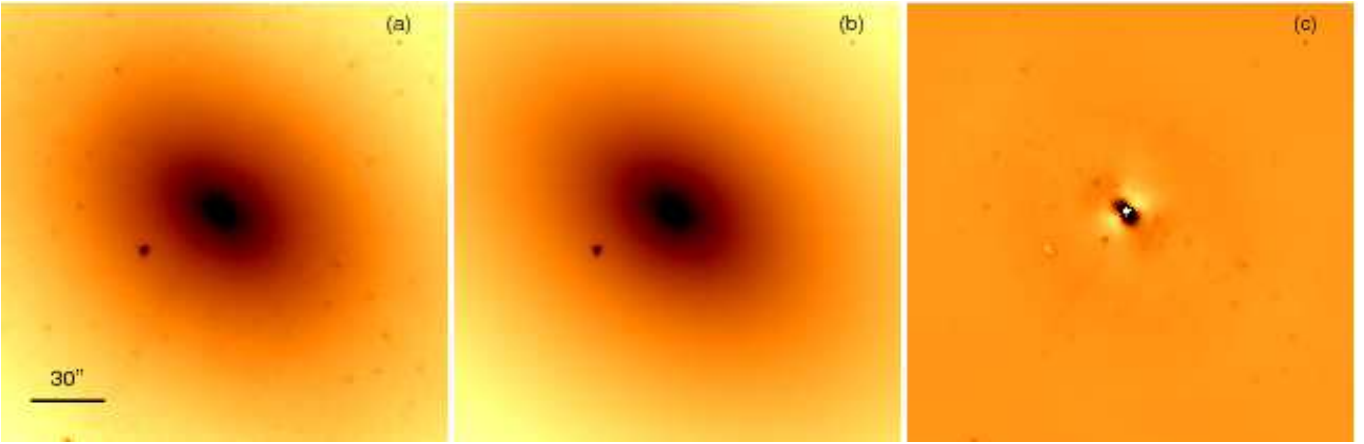}}
\caption{The input image used in GALFIT when fitting 
3.6\um~(a), the final model combining a \sersic profile, 
a point source, and sky (b), and the fit residuals (c). The 
4.5\um~images are similar to the 3.6\um~images, and are 
therefore not shown here. Each image is 
$3\farcm0\times 3\farcm0$ with north to the top and east 
to the left, as are all images throughout this paper 
unless stated otherwise. The color scale for all the 
panels is inverted so bright regions in the residual 
image (c) are regions where the model is 
over-subtracted. Panels a and b have logarithmic 
color scales, while panel c has a linear scale. 
The dark point source 35$\arcsec$ southeast of the galaxy's 
nucleus is a foreground star, which we also fit. The scale bar corresponds 
to 30$\arcsec$ (3.3~kpc).
\label{ch1}}
\end{figure*}

\begin{figure*}
\centerline{\includegraphics[width=\linewidth]{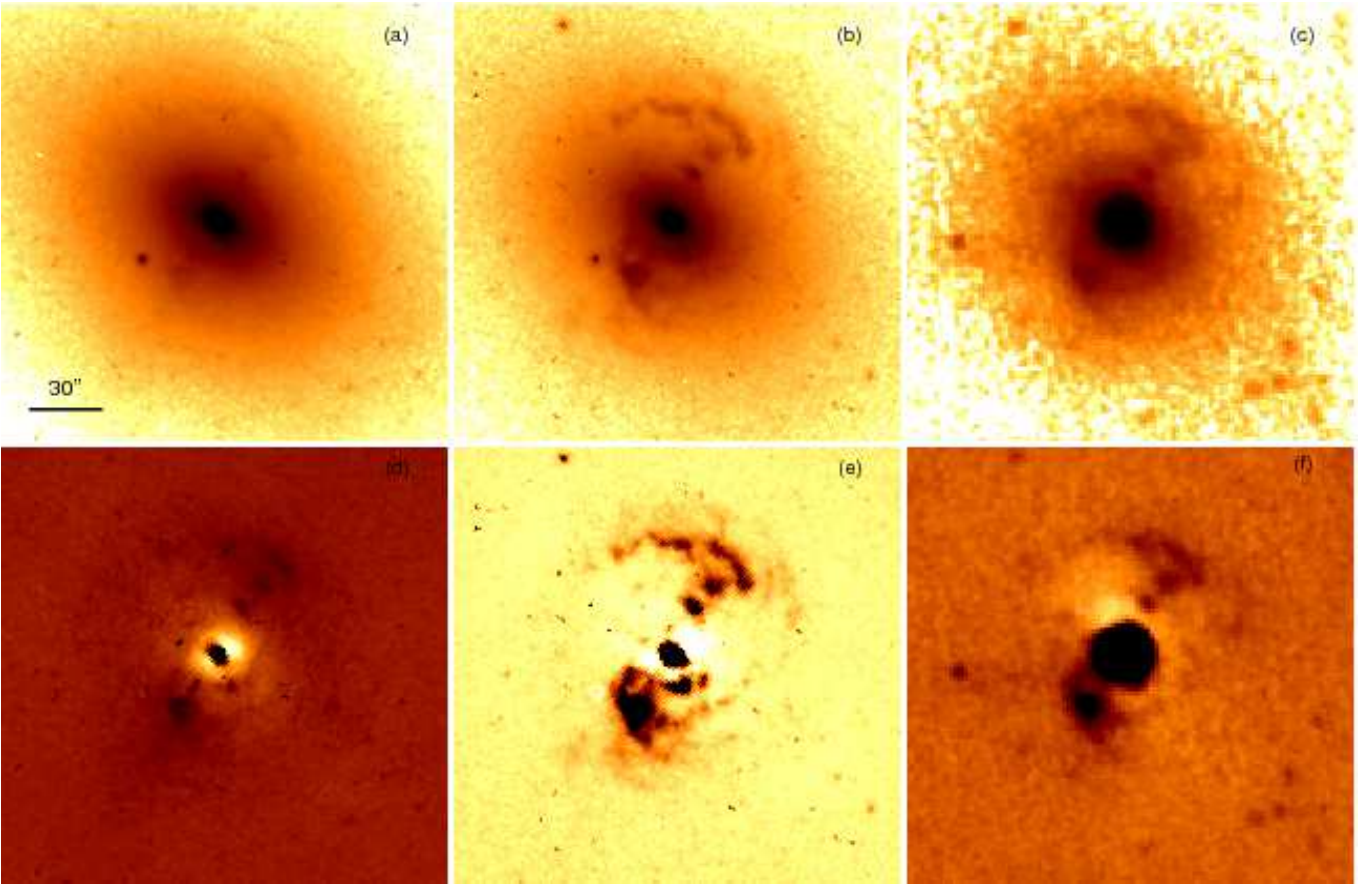}}
\caption{The top row shows the input image used in 
GALFIT when fitting the 5.8\um~(left), 8.0\um~(middle), 
and 24\um~(right) images. The bottom row are those same images, 
once a \sersic model and the foreground star have 
been subtracted. The image sizes are the same as in 
Figure \ref{ch1}. The color scale is inverted, and 
the top row has a logarithmic color scale, while the 
bottom row has a linear color scale. The dark regions 
in the residual images are excess emission, most likely 
due to dust. The scale bar corresponds to 30$\arcsec$ (3.3~kpc).
\label{ch345}}
\end{figure*}

A $12\arcmin\times12\arcmin$ region centered on the galaxy was selected for 
analysis in each IRAC and MIPS mosaic. The resulting 3.6\um~image, which is very
similar to the 4.5\um~image, is presented in Figure \ref{ch1}a. The 
corresponding images for 5.8\um, 8.0\um, and 24\um~are shown in the 
top row of Figure \ref{ch345}. Uncertainty maps were created by adding 
pixel-pixel rms noise and shot noise in quadrature. 

\subsection{\chandra~Observations}

NGC 1316 was observed for 30~ks on 2001 April 17 (ObsID 2022) 
with the \chandra~Advanced CCD Imaging Spectrometer 
Spectroscopy Array \citep[ACIS-S;][]{wei00}. This observation
was previously studied by \citet{kim03}. We reprocessed this 
observation applying the latest CTI and time-dependent gain 
calibrations (see Vikhlinin et al. 2005 for more details). 
We applied the standard filtering by grade, excluded bad/hot 
pixels and columns, and removed cosmic ray `afterglows'. We 
removed time intervals with background flaring, resulting in 
an effective exposure time of 23.853~ks. The background files 
(see Markevitch et al. 2000 for details) were processed in 
exactly the same manner as the observations.

\subsection{\xmm~Observations}
\xmm~observed NGC 1316 for 107.3~ks on 2005 August 11-12 
(ObsID 0302780101). Here we examine the Metal Oxide 
Semi-conductor CCD \citep[MOS;][]{jan01} observation. We 
filtered the data to remove periods of background flaring, 
resulting in a reduced exposure time of 62.0~ks for MOS1 and 
56.3~ks for MOS2. The events were further filtered to 
retain only events with energies between 0.5 and 7.0 keV 
and patterns between 0 and 12.

\subsection{Radio, Optical, and CO Observations}
NGC 1316 was observed with the $Very~Large~Array$ 
($VLA$) at 4.89~GHz in the AB array configuration on 
2002 June 1. We used the Astronomical Image Processing 
System (AIPS; version 31DEC09) package to generate a 
map with a resolution  of $1\farcs38\times1\farcs02$ 
(Figure \ref{xrI}c).  We also obtained a 20~cm \vla~map 
from the NASA/IPAC Extragalactic Database (NED) \citep{fom89}. 

The \hst~$Space~Telescope$ Advanced Camera for Surveys 
(ACS) observed NGC 1316 on 2003 March 04 through the F555W filter for
6.98~ks (PropID 9409). The WFPC2 instrument on \hst~observed it through
the F814W filter on 1996 April 07 for 1.86~ks (PropID 5990).
$^{12}$CO(2-1) intensities at 230 GHz were obtained
from C. Horellou and J.~H. Black, who observed NGC 1316 with
the 15 m Swedish-ESO Submillimeter Telescope (SEST)
in 1999 and 2001 with a resolution of 22$\arcsec$ \citep{hor01}.

\section{Data Analysis}

\subsection{\spitzer~Analysis}
To measure dust and nuclear emission, we first had to remove 
the stellar component from the \spitzer~images. To accomplish 
this, we modeled the 3.6\um~emission, where the flux from the 
stars is the greatest of all four IRAC bands, with a 
two-dimensional \sersic model \citep{ser68} for the stellar 
contribution. To determine the galaxy model, as well as emission 
from the nucleus, we used the software package GALFIT \citep{pen02}, 
which is a parametric surface brightness fitting code 
using $\chi^{2}$ minimization. We iteratively determined
the center for the \sersic profiles based on the
3.6\um~emission and the position of the central point source
based on the 8\um~emission. After confirming that the 
4.5\um~emission results in similar parameters, as expected, 
since this band is also dominated by stellar emission, 
we held the \sersic index, effective radius, axis 
ratio, position angle, and central position fixed at the 
3.6\um~fitted values for subsequent fits to the 4.5\um, 
5.8\um, 8.0\um, and 24\um~images. The remaining free parameters 
were the normalizations of the \sersic model, central point 
source, and the foreground star 35$\arcsec$ southeast of the nucleus, 
as well as a sky model, consisting of a constant offset 
and gradients in the two array directions. The point response 
function (PRF) in each band was input to GALFIT in order 
to fit the emission from the nucleus and foreground star. 
Masks were used to exclude foreground stars and the bulk 
of the emission from the neighboring galaxy NGC 1317.  
The fitted model parameters are given in Table \ref{fitTable}. 
The fitted \sersic index of $6.07\pm0.10$ agrees well 
with indices of 5.8 and 5.9 determined by \citet{cot07} 
from \hst~ACS images of NGC 1316.

Figure \ref{ch1}a shows the 3.6\um~mosaic. Figure 
\ref{ch1}b shows the best fitting model obtained from 
GALFIT of a central point source, foreground point source,
and a \sersic model. Since the residual emission, shown in 
Figure \ref{ch1}c, is on the order of the noise ($\sim3\%$), 
outside the nuclear region, our \sersic model is a good 
stellar emission model for this galaxy. Since the 4.5\um~image
and residuals are similar to those at 3.6\um, they are not shown. 

Figure \ref{ch345} presents the 5.8\um, 8.0\um, and 24\um~images 
(top row) and residuals after the modeled \sersic profile has 
been subtracted (bottom row). The 8.0\um~image
and contours of its nonstellar emission were previously shown
in \citet{tem05}. The non-axially symmetric, non-stellar 
component is visible in the top row (a-c) and quite 
striking in the bottom row (d-f). As we discuss in \S4 and 
\S5.1, most of this component is due to 
dust. In addition to the nucleus, there is a 
region approximately 11\farcs8 (1.3~kpc) in radius 
of dust emission 29$\arcsec$ (3.1~kpc) southeast of the nucleus. There 
is also dust emission extending 44$\arcsec$ (4.8~kpc) 
towards the northwest, ending in a clumpy arc that extends over a 
$\sim$90\dg~angle and containing two knots 21\farcs8 (2.4~kpc) 
and 32\farcs0 (3.5~kpc) from the nucleus. 

\begin{deluxetable*}{lccccc}
\tabletypesize{\scriptsize}
\tablecaption{GALFIT determined parameters for \sersic 
   model and nuclear point source\label{fitTable}}
\tablewidth{0pt}
\tablehead{
\colhead{Band} & \colhead{\sersic Integrated Flux}   & \colhead{Sky} 
     & \colhead{Nuclear Flux} & \colhead{Nuclear Luminosity}\\
\colhead{($\mu$m)}  & \colhead{(mJy)}  & \colhead{DC Offset}
     & \colhead{(mJy)}& \colhead{($10^{41}$~\ergss)}
}
\startdata
3.6   & $2390\pm70$  & -0.10 &  $4.67\pm0.22$ &  $2.40\pm0.40$\\
4.5   & $1420\pm40$  & 0.57  &  $5.63\pm0.26$ &  $2.32\pm0.38$\\ 
5.8   & $1190\pm40$  & -0.05 &  $6.78\pm0.27$ &  $2.16\pm0.35$\\ 
8.0   & $560\pm17$   &  0.11 &  $16.6\pm0.5$  &  $3.85\pm0.62$\\ 
24    & $400\pm12$   & -0.20 &  $60.8\pm1.8$  &  $4.69\pm0.76$\\ 
\enddata
\tablecomments{The \sersic model index ($n = 6.07\pm0.10$), effective 
radius ($r_{eff}=146\farcs2\pm0\farcs6$), axis ratio ($0.688\pm0.004$), 
and position angle (234.0\dg$\pm0.2$\dg~east of north) were 
all fitted only at 3.6\um~and were held constant for the fits 
made in the other bands. The central position of the
\sersic~profiles and the position of the nucleus were determined
iteratively using the 3.6\um~and 8\um~images to be
($03^{h}22^{m}41^{s}.69$, -37\dg12$\arcmin$28\farcs8) and
($03^{h}22^{m}41^{s}.71$, -37\dg12$\arcmin$28\farcs7) respectively.
Note that the \sersic index determined matches well with 
the ones obtained by C\^{o}t\'{e} et al. (2007).  Luminosities in 
each photometric band given here 
are $\nu L_{\nu}$ in\ergss, determined from the
luminosity density $L_{\nu}$. The uncertainty in the
luminosities is the result of both the uncertainty
in the flux measurement and in the distance to NGC 1316.}
\end{deluxetable*}

\subsection{\chandra~Analysis}
\begin{figure*}
\centerline{\includegraphics[width=\linewidth]{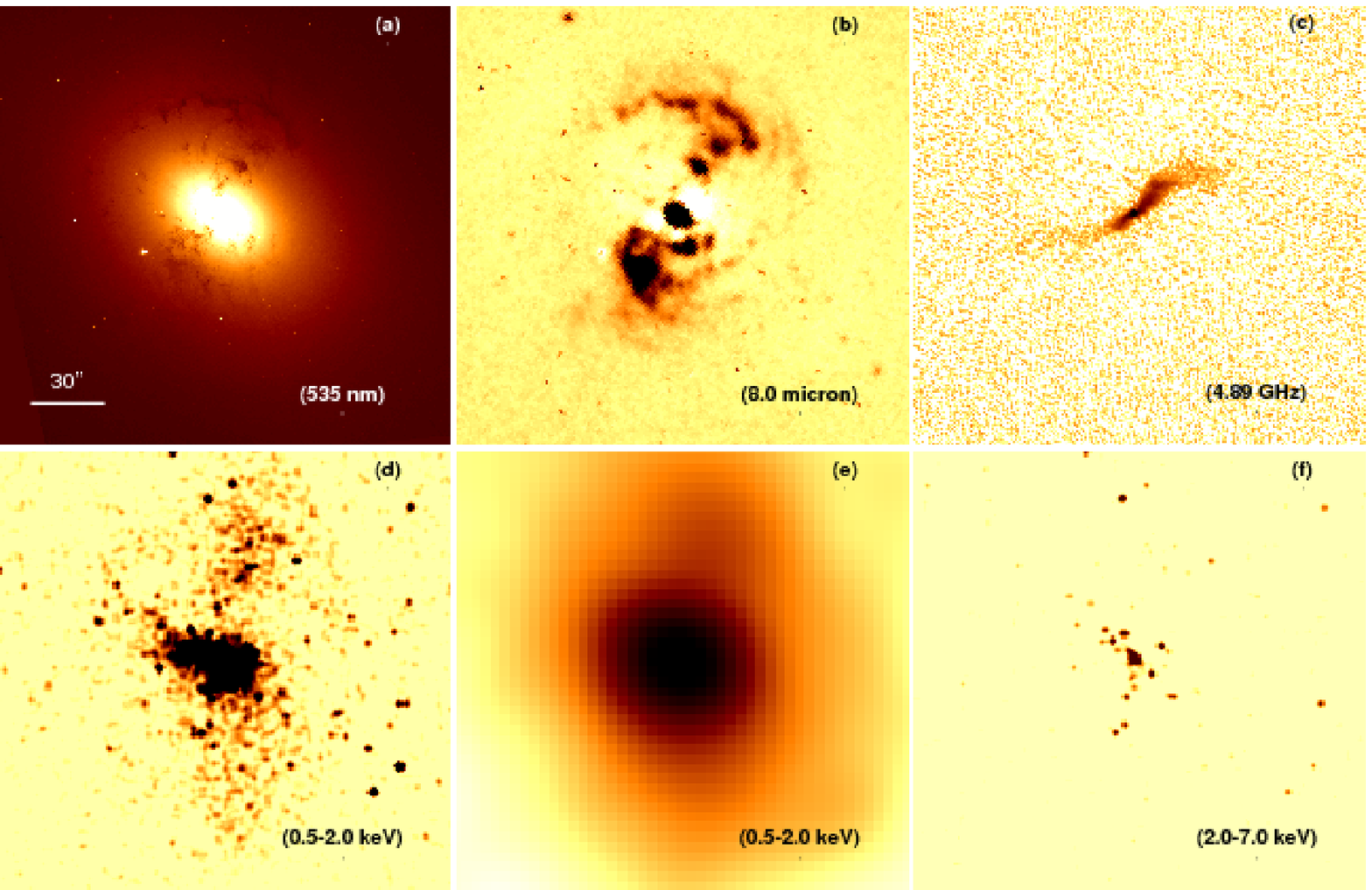}}
\caption{$3\farcm0\times3\farcm0$ images of NGC 1316: 
(a) \hst~ACS visible emission (535 nm), 
(b) \spitzer~non-stellar 8.0\um~emission, 
(c) \vla~4.89~GHz emission with a resolution of $1\farcs38\times1\farcs02$, 
(d) \chandra~soft X-ray (0.5-2.0~keV) emission,
(e) \xmm~soft X-ray (0.5-2.0~keV) emission,
(f) \chandra~hard X-ray (2.0-7.0~keV) emission.  The \chandra~images 
are smoothed with a 5$\arcsec$ Gaussian and have pixel sizes
of 1$\arcsec$ and 0\farcs5 for the soft and hard images respectively. The 
\xmm~image is the central region of Figure \ref{xmm}, demonstrating the
north-south elongation of soft X-ray emission.
With the exception of panel a, the darker regions have more emission. 
Panels a, c, and e have a logarithmic color scale, while panels b, d, and f have
a linear color scale. The scale corresponds to 
30$\arcsec$ (3.3~kpc). These images show the variety of morphology present
in this galaxy. Optical dust extinction coincides with infrared
dust emission, but the distribution of the dust is distinctly different
from that of the hot X-ray emitting gas. 
\label{xrI}}
\end{figure*}

We analyzed \chandra~observations of NGC 1316 to determine 
the brightness and morphology of emission due to hot gas 
and a central nuclear source. We created an image of the 
soft X-ray emission between 0.5 and 2~keV using the ACIS-S3 CCD. 
We removed point sources, which we detected using WAVDETECT 
\citep{fre02}. We then created an exposure map that 
accounts for all position dependent, but energy independent, 
efficiency variations across the focal plane (e.g., overall 
chip geometry, dead pixels or rows, variation of telescope 
pointing direction). Finally, we made a flat-fielded image 
by subtracting both the blank-field background and the read-out background 
and then dividing by the exposure map. We also created an image of the hard 
X-ray emission between 2 and 7~keV. The
resulting images, shown in Figures \ref{xrI}d and \ref{xrI}f, 
have 1$\arcsec$ and 0\farcs5 pixels respectively and have been 
smoothed with a 5$\arcsec$ Gaussian.

\subsection{\xmm~Analysis}
\begin{figure*}
\centerline{\includegraphics[width=0.75\linewidth, keepaspectratio]{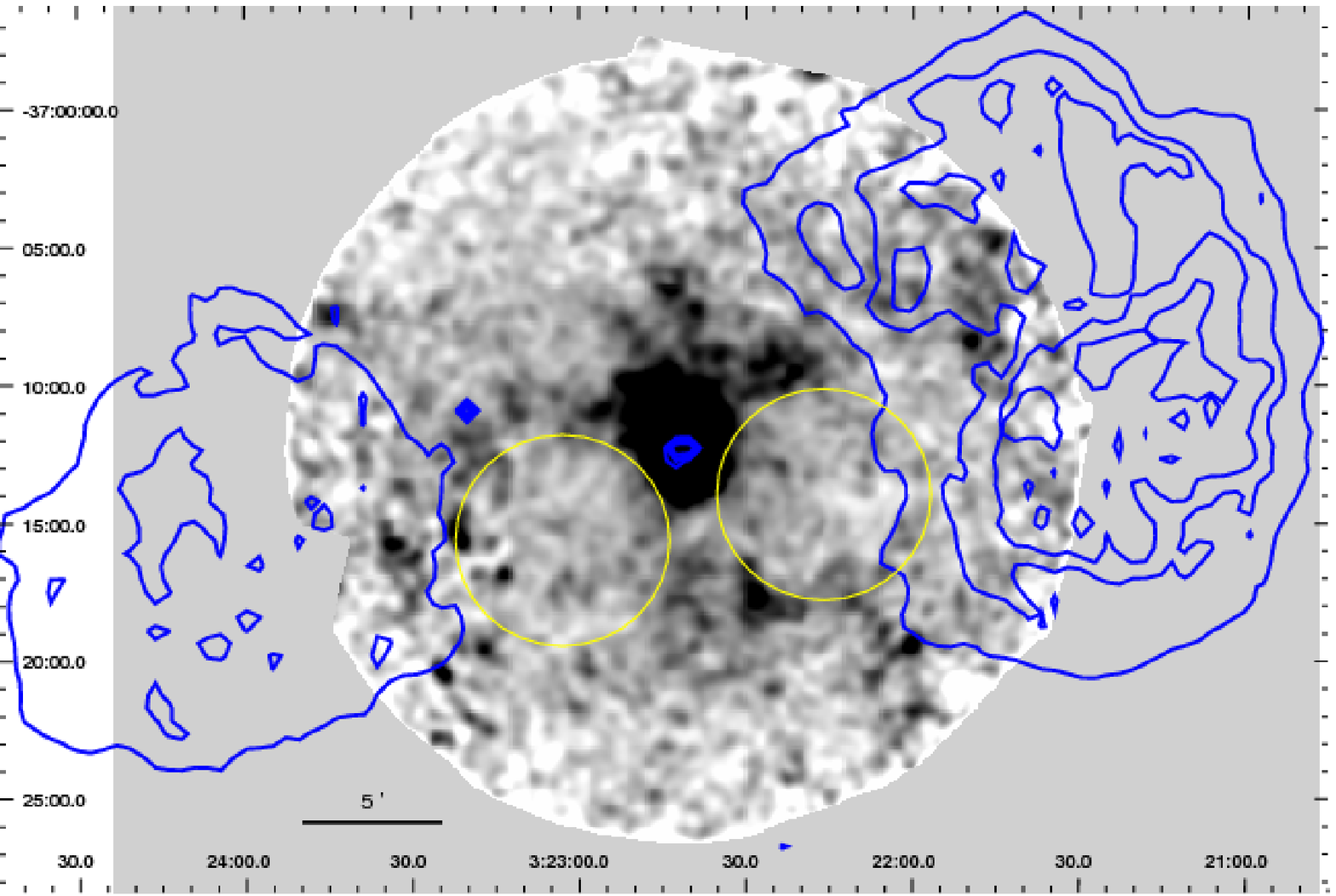}}
\caption{$30'$ soft (0.5-2.0~keV)  exposure-corrected \xmm~MOS 
image. Large X-ray cavities (shown by yellow 
circles) can be seen to the west (right) and to the southeast as 
lighter regions, centered at ($3^{h}22^{m}16^{s}$,~-37\dg14$\arcmin$00$\arcsec$) and
($3^{h}23^{m}3^{s}$,~-37\dg15$\arcmin$40$\arcsec$) respectively. The edges 
of the western cavity are marked by emission 
at ($3^{h}22^{m}28.5^{s}$,~-37\dg17$\arcmin$42$\arcsec$) and 
($3^{h}22^{m}24^{s}$,~-37\dg09$\arcmin$35$\arcsec$). 
The southeastern cavity has a faint edge to its north at 
($3^{h}23^{m}00^{s}$,~-37\dg11$\arcmin$33$\arcsec$).
These cavities contrast to the brighter emission to the north 
and south of the nucleus. The 
blue contours at 6, 11, 15, and 20 mJy/beam are from a 20~cm radio 
image and show the location of the radio lobes. 
The scale bar corresponds to 5$\arcmin$ (32.6~kpc). 
\label{xmm}}
\end{figure*}

We analyzed the \xmm~image to examine features of
the extended X-ray emission which were not in the smaller \chandra~field of 
view. We created new exposure maps and images, to remove 
the contribution from the MOS1 CCD with higher background, and 
we removed point sources. The extended 
X-ray emission, described in \S 4.3, includes a pair of cavities 320$\arcsec$
(34.8~kpc) to the west and southeast of the nucleus. We reprojected 
blank sky background files for each of the MOS 
CCDs\footnote{http://xmm.vilspa.esa.es/external/xmm\_sw\_cal/ \\
background/blank\_sky.shtml}
to the coordinates of our event file and scaled them appropriately,
prior to subtracting the combined background file from the
image. We then divided by the exposure map to create an 
exposure-corrected image with 4$\arcsec$ pixels (Figure \ref{xmm}).

\begin{figure*}
\centerline{\includegraphics[width=0.55\linewidth, keepaspectratio]{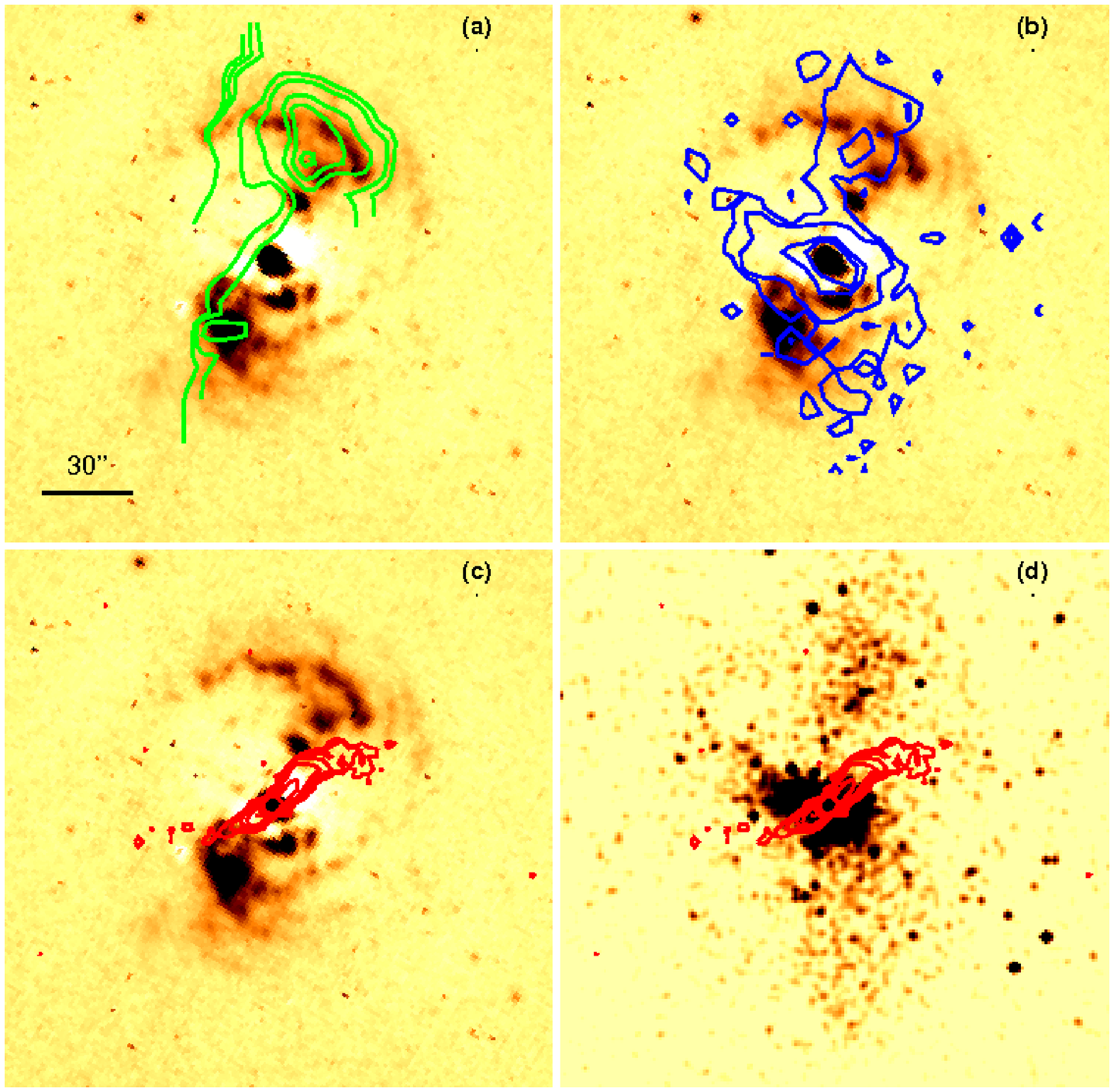}}
\caption{$3\farcm0\times3\farcm0$ images of NGC 1316: 
(a, b, c) \spitzer~non-stellar 8.0\um~emission with (a) 
contours of CO(2-1) (green) \citep{hor01}, (b)
flat-fielded soft (0.5-2.0~keV) \chandra~X-ray contours (blue) 
at 0.01, 0.02, 0.1, and 0.2~counts/ks/arcsec$^{2}$, and (c)
radio contours (red) at 0.04, 0.06, 0.10, 0.20, 0.50, 
1.0,  and 6.0~mJy/beam (rms is 0.03 mJy/beam); and
(d) \chandra~soft X-ray (0.5-2.0~keV) emission with 
radio contours. The scale corresponds to 
30$\arcsec$ (3.3~kpc). These images show that: 1) infrared dust and molecular
emission coincide, 2) the soft X-ray emission shows no indication
of absorption due to the dust and cold gas, and 3) that the resolved radio 
jet does not coincide with dust or soft X-ray emission, but lies 
in an inner X-ray cavity.
\label{xrIcon}}
\end{figure*}

\section{Multiwavelength Comparison of Features}

Figure \ref{xrI} shows NGC 1316 in visible (555 nm), non-stellar infrared (8.0\um), 
radio (4.89~GHz), \chandra~soft (0.5-2.0~keV), \xmm~soft (0.5-2.0~keV), 
and \chandra~hard (2.0-7.0~keV) emission. For the 8.0\um~image, we
subtracted the stellar emission, as modeled by a \sersic profile.
In Figure \ref{xrIcon}, we show the relative locations of the
X-ray, IR, radio, and CO emission. Below, we discuss the nucleus, 
the jet, and the extended non-stellar emission. 

\subsection{Nuclear Emission}
The nucleus is detected in the radio, UV, X-ray, and IR bands. 
We obtained radio and UV nuclear fluxes from the literature 
\citep{gel84, fab94}. 
We measured the X-ray spectrum in a 1$\arcsec$ radius circle around
the nucleus, using a 1$\arcsec$-2$\arcsec$ annulus to subtract the
background thermal emission. We fit the nuclear spectra with an
absorbed power law with both a variable $n_{H}$ and an $n_{H}$ 
held constant at the Galactic value of $2.4\times 10^{20}$~cm$^{-2}$ 
\citep{kal05}. The variable $n_{H}$ fit had a similar power law
index to the constant $n_{H}$ fit and a column density not significantly
larger than the Galactic absorption. Therefore, we used the constant $n_{H}$
spectral fit whose best-fit power law index was $2.09^{+0.41}_{-0.46}$ 
(90\% confidence), which is within the range 
typically observed for active galactic nuclei (AGN). 
 We converted measured soft (0.5-2.0~keV) and hard 
(2.0-7.0~keV) X-ray fluxes to luminosities of 
$6.5 \pm 1.8 \times 10^{38}$~\ergss~and $5.7 \pm 2.1 \times 10^{38}$~\ergss. 
We measured the X-ray nuclear broadband (0.3-8~keV) flux to be  
$2.4 \pm 0.7 \times 10^{-14}$~\ergsscm. Our flux corresponds to a 
broadband (0.3-8.0~keV) luminosity of $1.5 \pm 0.5 \times 10^{39}$~\ergss. 
We also measured the nuclear flux in the F814W \hst~image in a $0\farcs15$ 
radius aperture to be $7.8 \pm 0.2 \times 10^{-13}$~\ergsscm, which we consider
an upper limit because of the stellar contribution. Since the F555W image 
is saturated in the central region, we do not measure the nuclear flux 
through this filter.

\begin{figure*}
\centerline{\includegraphics[width=\linewidth]{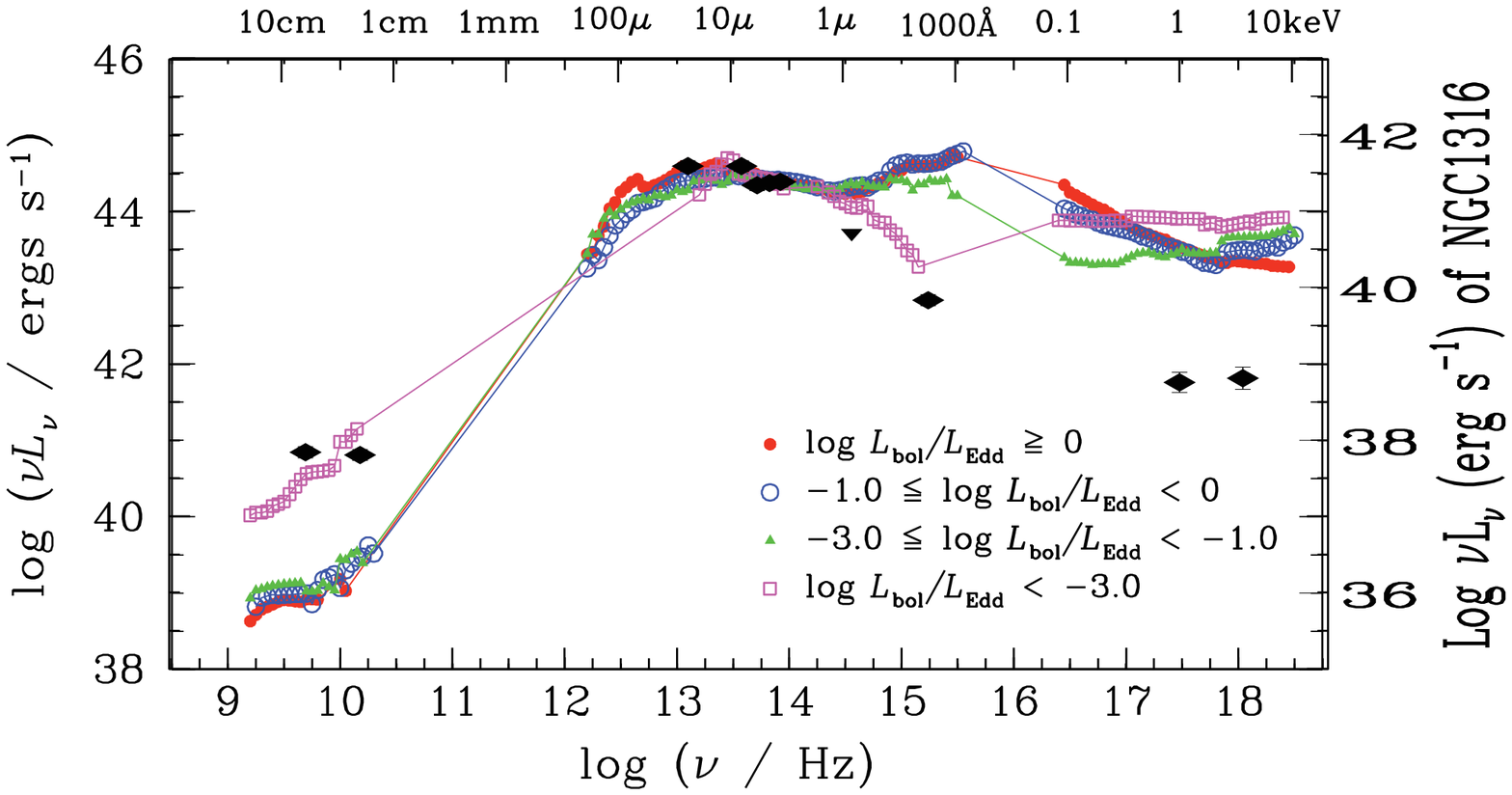}}
\caption{Spectral energy distribution of the nuclear emission
in the \spitzer~and \chandra~bandpasses, through the F814W \hst~filter (upper 
limit), at 4.9~GHz and 15.0~GHz from \citet{gel84}, and at 1730\AA~from 
\citet{fab94}. The right vertical axis applies to this AGN, and the
left axis to the comparison AGN SEDs that are normalized at 1\um~and
come from Figure 7 of \citet{ho08}. The purple squares are the LLAGN
with log($L_{bol}/L_{Edd} < -3.0$). NGC 1316's AGN has
log($L_{bol}/L_{Edd}= -3.9$). Note the large radio to optical ratio and
the lack of the UV-optical big blue bump in the NGC 1316 AGN.
\label{sed}}
\end{figure*}

The nucleus is detected in all five \spitzer~bands. Fluxes and
luminosities, $\nu L_{\nu}$, in each photometric band from the GALFIT model
are given in Table \ref{fitTable}. We note that the nuclear fluxes 
were derived from fitting, but that the point source is not 
directly visible in the IRAC images. To confirm that the galaxy 
does contain a point-like nucleus, we used the 3.6\um~emission as a 
stellar model, which was then color corrected and scaled to correct
for differences in the apertures and zero point magnitudes between the IRAC
bands. The resulting non-stellar images are very similar to those in the bottom
row of Figure \ref{ch345}. The IR color of NGC 1316's nucleus falls outside, but 
within $3 \sigma$, of the region defined by
the mid-IR AGN color selection criteria of \citet{ste05} and 
within the selection region defined by \citet{lac04}. We note however that these
IRAC selection criteria were developed for Seyfert galaxies and quasars that
are significantly more luminous than NGC 1316. Indeed, the spectral
energy distribution (SED) of the NGC 1316 AGN (Figure \ref{sed}) 
is similar to those of other low luminosity AGN (LLAGN) 
\citep{ho99} in that it lacks the big blue bump found
in powerful, optically bright AGN and instead appears to show only 
a single big red bump and has a larger radio to optical ratio than that of the 
higher luminosity AGN \citep{ho08}. The comparison spectra
plotted in this figure are from Figure 7 of \citet{ho08} where
the purple squares are the LLAGN. \citet{smi07} cited
the AGN of NGC 1316 as having the typical peculiar PAH spectrum
of LLAGN, which has low ratios of $L$(7.7\um)/$L$(11.3\um).

We used the supermassive black hole mass of 
$1.5\times 10^{8}$~\msun~\citep{now08} to determine the Eddington 
luminosity for the AGN to be $2.3\times 10^{46}$\ergss.
We interpolated the SED between the observed photometric points from 
10$^{9}$ Hz to 10$^{18}$ Hz and derived a bolometric 
luminosity of $\sim2.4\times 10^{42}$\ergss,
corresponding to a bolometric correction of $\sim$6.2 for the 8.0\um~IRAC band.
The AGN therefore has a low Eddington ratio of $\sim10^{-4}$. We also calculated the 
Bondi accretion rate to be $1.6\times10^{-4}$~\msun~yr$^{-1}$ 
\citep{bon52}, based on the black hole mass \citep{now08}, 
the 0.77 keV gas temperature \citep{iso06},
and a central gas density of 0.4~cm$^{-3}$ derived from a  
$\beta$ model \citep{cav76} fit to the central 200$\arcsec$ (21.7~kpc) region using \chandra~data
(a$_{X}=$~320~pc, $\beta=0.49$, $n_{0}=0.4$~cm$^{-3}$) \citep{jon10}.

\begin{deluxetable}{cccccc}
\tabletypesize{\scriptsize}
\tablecaption{Jet Infrared Flux Upper Limits \label{jetTable}}
\tablewidth{0pt}
\tablehead{
\colhead{Band} & \colhead{Flux} & \colhead{1$\sigma$ Unc.} & 
   \colhead{Luminosity} & \colhead{1$\sigma$ Unc.} & \\
\colhead{($\mu$m)} & \colhead{(mJy)} & \colhead{(mJy)} &
   \colhead{($10^{40}$~\ergss)} & \colhead{($10^{40}$~\ergss)}
}
\startdata
3.6  & $<$1.05  & 0.35 &  $<$5.94  & 1.98 \\
4.5  & $<$0.66  & 0.22 &  $<$2.98  & 0.99 \\
5.8  & $<$0.58  & 0.19 &  $<$2.07  & 0.69 \\
8.0  & $<$0.39  & 0.13 &  $<$0.97  & 0.32 \\ 
24   & $<$0.81  & 0.27 &  $<$0.69  & 0.23 \\ 
\enddata
\tablecomments{Fluxes, given in mJy, are 3$\sigma$ upper limits
and were obtained using an aperture defined by the radio emission
northwest of the nucleus applied to each \spitzer~image. The 
galaxy flux was removed using the \sersic model determined 
from fitting the \spitzer~images with GALFIT. Luminosities 
in each photometric band given here are $\nu L_{\nu}$ in\ergss, 
determined from the luminosity density $L_{\nu}$.}
\end{deluxetable}

\subsection{Inner Jet}
The 4.89~GHz radio emission on small scales (Figure \ref{xrI}c) 
has previously been extensively described by \citet{gel84}. The northwest jet
extends 0\farcm5 (3.3~kpc) from the nucleus and does not
decrease significantly in brightness over the initial 15$\arcsec$ (1.6~kpc). 
In contrast, the weaker southeast counterjet decreases
immediately in brightness away from the nucleus.
Unlike the jet in M87 
\citep{shi07, for07} which is clearly detected 
in all four IRAC bands, the NGC 1316 radio jet is not 
detected in any IRAC band or at 24\um. We used an aperture 
defined by the region of radio emission to the northwest of 
the nucleus to derive upper limits on the IR emission from the
jet, which are given in Table \ref{jetTable}.
Extended aperture corrections were derived and 
applied.\footnote{IRAC: Extended Source Calibration \\
http://ssc.spitzer.caltech.edu/irac/iracinstrumenthandbook/33} 
We also calculated the expected fluxes assuming
a synchrotron model with a typical spectral index of 0.55 from
the radio flux of 29 mJy within the aperture. The expected
fluxes range from 0.38 mJy to 0.14 mJy, which are below
the limits measurable with the present data.

The soft X-ray image (Figure \ref{xrI}d and \ref{xrI}e
and contours on Figure \ref{xrIcon}b) does not show emission 
from the jet. The resolved northwest radio jet coincides 
with a region of low X-ray emission, most likely a small 
$\sim15\arcsec$ X-ray cavity created by the expansion 
of the radio plasma, previously described by \citet{kim03}.
As illustrated in Figure \ref{xrIcon}c, the dust emission 
is faint at the position of the radio jet. The bend in the 
northwestern jet is located just south of the first IR
knot, along the northwestern dust protrusion.  

\subsection{Extended Non-Stellar Emission}

\citet{tem05} found that the morphology of the 8.0\um~non-stellar 
emission was similar to that of the 15\um~emission
detected by the \emph{Infrared Space Observatory} (ISO) \citep{xil04}. They 
concluded that, while much of the excess emission
at 8.0\um~was likely due to PAH emission at 7.7\um,
warm, small dust grains also contributed. The 
similarity of the features at 5.8\um~and 24\um~supports this
interpretation. The extended non-stellar
emission has significant structure at all wavelengths as
shown in Figure \ref{xrI} and Figure \ref{xrIcon}:
\begin{itemize}
\item The \hst~ACS image (Figure \ref{xrI}a) demonstrates
  that regions of visible dust extinction have similar
  morphology to the infrared dust emission described in 
  \S 3.1 and shown at 8.0\um~in Figure \ref{xrI}b. (See
  similar image in \citet{tem05}.)
\item The CO contours \citep{hor01} superimposed
  on the 8.0\um~non-stellar emission in Figure \ref{xrIcon}a demonstrate that
  the northwestern and southeastern dust emission regions coincide 
  with molecular hydrogen traced by the CO emission,  
  suggesting a common origin for the dust and cold gas. 
\item The strongest X-ray emission outside the nucleus 
  (Figure \ref{xrIcon}b with soft X-ray contours 
  overlaid on the 8.0\um~non-stellar emission)
  extends northeast of the nucleus along the major axis of NGC 1316 into 
  a region absent of infrared dust emission. We tested whether
  the dusty features seen in the infrared and the coincident cold
  gas result in soft X-ray absorption, but
  found no indication thereof, although the data is not sufficient to
  make a conclusive statement. 
\end{itemize}

In the galaxy core, the soft \chandra~image and the \xmm~image 
show a roughly north-south elongation 
approximately 1\farcm25 (8.2~kpc) in each direction (Figures 
\ref{xrI}d, \ref{xrI}e), which does not follow the
distribution of the stars. Instead, this emission is roughly
perpendicular to the major axis of NGC 1316 and may be from hot gas
that was moved by the outburst.  The larger \xmm~field 
of view (Figure \ref{xmm}) shows further filamentary emission
 north of the nucleus and a pair of X-ray cavities. 
These cavities, likely created by the expansion of radio 
plasma, are marked with yellow circles of 230$\arcsec$ (25~kpc) radii. 
The western cavity is centered at ($3^{h}22^{m}16^{s}$,~-37\dg14$\arcmin$00$\arcsec$)
and the southeastern cavity is centered at 
($3^{h}23^{m}3^{s}$,~-37\dg15$\arcmin$40$\arcsec$). Each cavity 
lies 320$\arcsec$ (34.8~kpc in the plane of the sky) from the nucleus. 
There are three regions of 
enhanced emission along the edges of these cavities, approximately
located at  ($3^{h}22^{m}28.5^{s}$,~-37\dg17$\arcmin$42$\arcsec$),
($3^{h}22^{m}24^{s}$,~-37\dg09$\arcmin$35$\arcsec$), and 
($3^{h}23^{m}00^{s}$,~-37\dg11$\arcmin$33$\arcsec$), which are likely due
to increased gas density as the hot ISM is compressed by the 
expanding cavities. No radio emission is detected in these
X-ray cavities, a situation previously seen in Abell 4059 \citep{hei02}, M87 \citep{for07},
and the Perseus cluster \citep{fab06}. While the centers of the radio 
lobes line up with the AGN, there are indications that this system
may be experiencing some sloshing of the hot gas. Specifically, the 
X-ray cavities are centered 1\farcm5 and 3\farcm2 south
of the nucleus and \citet{eke83} found low-level radio emission 
between the lobes $\sim7\arcmin$ south of the nucleus.
\begin{figure}
\centerline{\includegraphics[width=\linewidth]{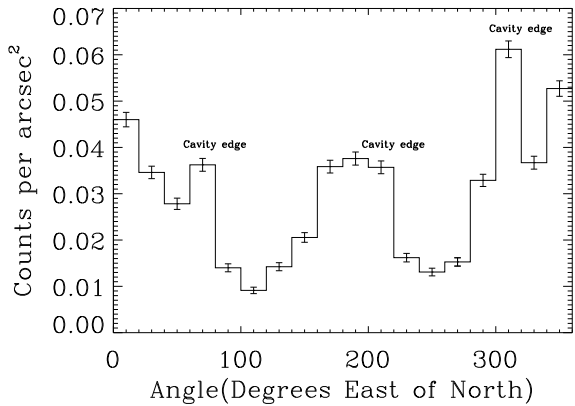}}
\caption{Azimuthal profile of background-subtracted counts per
square arcsecond for a radial sector with inner and outer radii of 180$\arcsec$ 
(19.6~kpc) and 375$\arcsec$ (40.8~kpc) from the nucleus in the soft (0.5-2.0 keV) 
\xmm~image in Figure \ref{xmm}. Angles are given counterclockwise 
from north. Note that there is 
less flux to the southeast (120\dg) and to the west (250\dg), 
 which corresponds to the locations of the centers of the cavities 
apparent in Figure \ref{xmm}.
\label{azprof}}
\end{figure}

To quantitatively measure the significance of the cavities seen in
the \xmm~image, we plot in Figure \ref{azprof} the azimuthal surface 
brightness of the soft (0.5-2.0~keV) exposure-corrected background-subtracted
\xmm~image (Figure \ref{xmm})
taken in an annulus extending from 180$\arcsec$ (19.6~kpc) to 
375$\arcsec$ (40.8~kpc) from the nucleus after the image was 
smoothed with a 28\farcs2 Gaussian. 
The azimuthal profile shows that the regions between
80-140\dg, (i.e. southeast of the nucleus), and between
220-280\dg, (i.e. to the west), have significantly lower surface 
brightness than the rest of the annulus. These regions 
coincide with the cavities identified (yellow circles) 
in Figure \ref{xmm}. The bright regions north of the nucleus and 
along the cavity edges in Figure \ref{xmm} coincide with the 
significantly brighter regions in the azimuthal plot 
(Figure \ref{azprof}). The three regions of enhanced X-ray 
emission along the cavity edges are noted in Figure \ref{azprof}. 

We tested whether the variations in azimuthal surface brightness 
could be the result of abundance or gas density variations. The 
maximum surface brightness change would require a factor of 
2.3 difference in elemental abundance (i.e. the lower 
surface brightness region would have an elemental abundance 
40\% that of the brighter regions). While such an abundance 
gradient would be relatively long lived
against diffusion, even if it proceeds as fast as predicted for heavy
ions in a fully ionized plasma \citep{sar88,spi56}, such a distribution 
of metals mimicking cavity structures seems particularly 
contrived. An alternative explanation for the surface brightness 
variations is for the isobaric gas to have a density in the 
regions of lower surface brightness 0.66 times that of the gas 
to the north and south of the nucleus. Such a difference in 
density requires either that the lower surface brightness gas 
is at least 1.5 times hotter than the surrounding gas or that 
the regions of lower surface brightness would be filled with 
a relativistic plasma. No indication of emission from a hot 
plasma is seen in the harder X-ray band (2.0-7.0~keV). Therefore, 
we expect the regions of lower surface brightness to be cavities 
filled with a currently undetected relativistic plasma, a 
morphology seen in other galaxies and galaxy clusters and known 
as ghost cavities \citep{hei02}. Emission from ghost cavities 
are generally detected in low-frequency radio data (e.g. 
Giacintucci et al. 2009).

\section{Multiwavelength View of the Consequences of the Merger Event}

NGC 1316 exhibits signs of a recent merger, including
nuclear activity and a disturbed morphology seen in the optical 
and infrared dust distribution as well as the tidal tails first 
noted by \citet{sch80}. Each wavelength provides different 
insights into the merger event and the resulting structure of 
NGC 1316. Below, we discuss the distribution of the infrared-emitting 
dust and estimate the mass of the galaxy that collided with NGC 1316 from
the measured dust mass. We also use the morphology of the large scale
radio and X-ray emission (Figure \ref{xmm}) to constrain the recent 
outburst history of the central SMBH.

\subsection{Dust Distribution}
\begin{figure}
\centerline{\includegraphics[width=\linewidth]{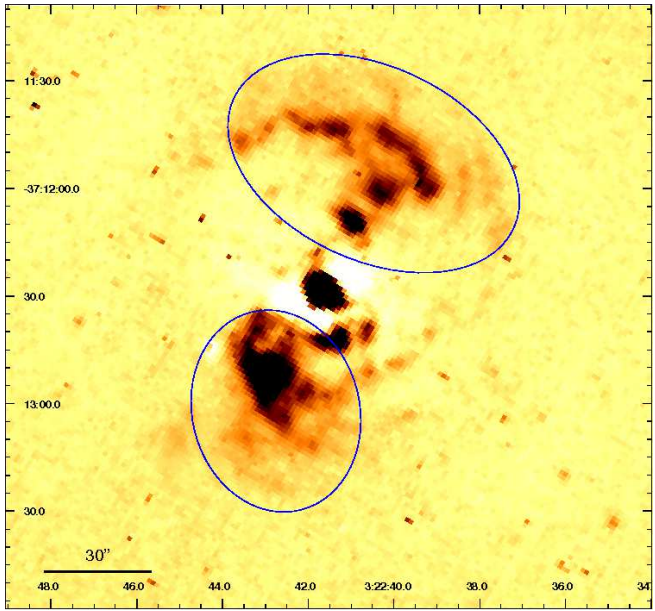}}
\caption{8.0\um~non-stellar image showing the position and size
of the apertures used to determine the dust photometry. The southeastern
aperture is an ellipse with a major axis of 56\farcs8, a minor 
axis of 46\farcs6, and a position angle of 284\dg~(east of north) centered at 
($3^{h}22^{m}42^{s}.75$,~-37\dg13$\arcmin$02\farcs1). The northwestern
aperture is an ellipse with a major axis of 86$\arcsec$, a 
minor axis of 54$\arcsec$, and a position angle of 245\dg centered at 
($3^{h}22^{m}40^{s}.47$,~-37\dg11$\arcmin$53\farcs0). The scale bar 
corresponds to 30$\arcsec$ (3.3~kpc).
\label{phot}}
\end{figure}

To measure the dust emission in the \spitzer~bands,  
we performed aperture photometry on the infrared 
images after subtracting a \sersic model of the stellar emission. Elliptical 
apertures (Figure \ref{phot}) were chosen 
to include dust features seen at 8.0\um. For the southeastern
region, we used an ellipse with a major axis of 56\farcs8, a minor 
axis of 46\farcs6, and a position angle of 284\dg~(east of north) centered at 
($3^{h}22^{m}42^{s}.75$,~-37\dg13$\arcmin$02\farcs1). For the 
northwestern region, we used an ellipse with a major axis of 86$\arcsec$, a 
minor axis of 54$\arcsec$, and a position angle of 245\dg~centered at 
($3^{h}22^{m}40^{s}.47$,~-37\dg11$\arcmin$53\farcs0).  
The counts in each aperture were background subtracted
and converted to fluxes, after an extended aperture correction was 
applied. Columns 2-3 of Table \ref{dustTable} list the flux in 
each aperture. The next two columns give the total fluxes and their 
uncertainties with and without the nuclear point source 
(nuclear fluxes given in Table \ref{fitTable}). The uncertainties 
have two components: the uncertainty in the \sersic model and the
uncertainty in the photometric accuracy of \spitzer~images \citep{rea05, eng07}. 

\begin{deluxetable}{lccccccc}
\tabletypesize{\scriptsize}
\tablecaption{Dust Photometry \label{dustTable}}
\tablewidth{\linewidth}
\tablehead{
\colhead{Band} & \colhead{SE Knot} & \colhead{NW Knot+Arc} &
    \colhead{Total} & \colhead{Total} &  \\
\colhead{} & \colhead{$03^{h}22^{m}42^{s}.75$} &
   \colhead{$03^{h}22^{m}40^{s}.47$}  & \colhead{of} &
   \colhead{with} & \\
\colhead{} & \colhead{-37\dg13$\arcmin$02\farcs1} &
   \colhead{-37\dg11$\arcmin$53\farcs0}  & \colhead{knots} &
   \colhead{nucleus\tablenotemark{1}} & \\
\colhead{($\mu$m)} & \colhead{(mJy)}  & \colhead{(mJy)} &
   \colhead{(mJy)} & \colhead{(mJy)}
}
\startdata
5.8 &  4.16$\pm$0.43 &  6.47$\pm$0.70 &   10.6$\pm$0.8 & 17.4$\pm$0.9  \\
8.0 & 18.8$\pm$0.6   & 13.8$\pm$0.5 & 32.6$\pm$0.8 & 49.2$\pm$0.9  \\ 
\enddata
\tablecomments{Fluxes (cols 2-3) were obtained using elliptical
apertures (Figure \ref{phot}) for the southeast and northwest 
region on each \spitzer~image once a \sersic model had been 
subtracted.}
\tablenotetext{1}{Nuclear fluxes are given in Table \ref{fitTable}}
\end{deluxetable}

\citet{dra07} modeled the integrated IRAC, MIPS, and IRAS 
fluxes of the SINGS galaxy sample with a two-component dust model 
and determined dust masses. They estimate their models are 
accurate to 10\%. To test whether the dust mass determined for NGC 1316 by 
\citet{dra07} is located within the regions visible in 
Figure \ref{phot}, we compared our photometry from the two
regions of dust emission and the nucleus to their dust 
model predictions. We convolved the predicted \citet{dra07} 
dust flux SED for NGC 1316 (their Figure 14) through the 
appropriate response functions to calculate
the expected fluxes in the \spitzer~band passes. 
Our 5.8\um~and 8.0\um~total fluxes for the dust regions
and nucleus of $17.4\pm0.9$~mJy and $49.2\pm0.9$~mJy
are consistent with the \citet{dra07} model values of
$13.0\pm1.3$~mJy and $45.4\pm4.5$~mJy, which are the 
emission for the entire galaxy. The agreement leads us 
to conclude that the dust mass estimated by 
\citet{dra07} is contained in the non-stellar 
IR emission regions described in \S3.1 and in the nuclear 
region.
                                      
\subsection{Dust Mass and Merger Progenitor Mass}

In the following, we show that the dust observed in NGC 1316
is not native to the galaxy and use the dust mass to estimate
the mass of the merger galaxy. The clumpy morphology 
of the dust is significantly different from the smooth
elliptical distribution of the stars, so the NGC 1316 stars 
could not have expelled the dust. In addition, \citet{tra09}
found for nearby ellipticals that most of the non-stellar 8.0\um~emission 
is confined to the nuclear region. This also demonstrates
that the morphology of the dust emission in NGC 1316 is unusual.
\citet{tem09} found a correlation between the K-band and 
24\um~luminosities of elliptical galaxies. NGC 1316 has 
a particularly large 24\um~luminosity for its K-band luminosity, 
about an order of magnitude greater than predicted by 
the \citet{tem09} correlation.
While \citet{tem09} did not find a correlation between 
K-band luminosity and either 70\um~or 160\um~luminosity, 
NGC 1316's integrated luminosities at these wavelengths of
$1.4\times 10^{43}$~\ergss~and $1.5\times 10^{43}$~\ergss~\citep{dal07} 
are also more than an order of magnitude greater than found for 
the galaxies in the \citet{tem09} sample. The large 
infrared luminosities of NGC 1316 demonstrate an external 
origin for the dusty emission.

In the following, we estimate the mass of dust in NGC 1316 
as well as the dust mass expected to be in an elliptical 
galaxy the size of NGC 1316. \citet{mun09} provided
a formula (A8) for calculating the dust mass of a galaxy 
from its 24\um, 70\um, and 160\um~fluxes and its distance. 
We calculated that NGC 1316 has a total dust mass of 
$2.4 \pm 0.9 \times 10^{7}$~\msun, using the integrated MIPS 
fluxes from \citet{dal07}. While emission from the dust is 
clearly observed, the large uncertainty on the dust mass
results from the uncertainties in the distance to NGC 1316 
($22.7\pm1.8$~Mpc) and in the integrated MIPS fluxes ($0.43\pm0.02$~Jy 
at 24\um, $5.44\pm0.40$~Jy at 70\um, and $12.61\pm1.78$~Jy at 160\um).
We revised the \citet{dra07} 
dust mass for NGC 1316, which was found on the basis of 
SED fitting, for our assumed distance of 22.7~Mpc to be 
$3.2 \times 10^{7}$~\msun, which is consistent with our dust mass. We 
used the fluxes for the sample of elliptical galaxies in \citet{tem09}, along 
with their B-V colors \citep{dev91} and the color-dependent 
mass-to-light ratios of \citet{bel03}, to calculate the stellar and dust 
masses of the sample. We found that elliptical galaxies typically have 
dust-to-stellar mass ratios between $0.7-5.3\times10^{-7}$. Using these 
ratios, we estimate that NGC 1316 with its stellar mass of 
$5.3\times10^{11}$~\msun~(based on B-V=0.87 \citep{dev91}, 
K = 5.587 \citep{jar03}, and the relations of \citet{bel03}) had an 
intrinsic dust mass of $0.4-3\times10^{5}$~\msun,~$\lesssim 1$\% of 
the measured dust mass. We conclude that 
nearly all of the dust currently present in NGC 1316 
was contributed by a merger galaxy.

We can constrain the galaxy type and the stellar and gas mass of the 
merger galaxy from its estimated dust mass of 
$2.4 \pm 0.9 \times 10^{7}$~\msun. The merger galaxy had to 
be a late type galaxy as its stellar mass, were it a typical 
elliptical, would have been roughly 200 times the present 
stellar mass of NGC 1316. We calculated the stellar 
masses of the spiral galaxies in the SINGS sample 
using NED colors and the color-dependent stellar 
$M/L$ ratios of \citet{bel03}. Using the dust masses from 
\citet{dra07}, we found dust-to-stellar mass 
ratios between $0.4-3.4\times10^{-3}$ for Sa-Sm galaxies with Sc 
galaxies having the largest ratios. From these ratios and the merger 
galaxy dust mass, we estimate that the merger galaxy had a stellar 
mass in the range of $1-6\times10^{10}$~\msun, approximately 
10\% of NGC 1316's current stellar mass. Assuming typical galaxy colors
\citep{tri00}, we calculated the corresponding $L_{B}$ to be $0.7-
2\times10^{10}$~\lsun~using the ratios of \citet{bel03}. 
We estimate the corresponding cold gas masses based on 
gas-mass-to-light ratios of \citet{bet03} to be 
$2-4\times10^{9}$~\msun. \citet{ken03} found an upper limit on the 
mass of neutral hydrogen in NGC 1316 of $5.5 \times 10^{8}$~\msun, 
and \citet{hor01} estimate that NGC 1316 has 
$7.4 \times 10^{8}$~\msun~of molecular hydrogen gas, resulting
in a total cold gas mass of less than $1.3\times10^{9}$~\msun. 
Since the merger galaxy's estimated cold gas mass of 
$2-4\times10^{9}$~\msun~is larger than NGC 1316's present cold gas 
mass, some cold gas may have been ionized due to mixing with hot 
gas or used in star formation in the merger process.

\subsection{Outburst Ages}

The morphologies of the large scale GHz radio emission and the large
X-ray cavities in NGC 1316 suggest two possible interpretations of the 
recent outburst history of the NGC 1316 AGN. Based on the relative
location of the X-ray cavities and the radio lobes, we can conclude 
that either the 1.4~GHz radio features do not fully define the
extent of the radio lobes created in conjunction with the X-ray cavities
in the course of a single outburst or there were at least two 
outbursts, one resulting in the radio lobes and a more recent one 
creating the X-ray cavities seen in the \xmm~image.  In the two outburst scenario,
we expect the X-ray cavities, which lie at a smaller radius, to result 
from the more recent outburst, as they would otherwise have been disrupted by 
the expanding radio lobes.

Cavity ages can be estimated by assuming that 
the bubbles that create them rise buoyantly in 
the gaseous atmosphere \citep[e.g.][]{chu01}. Assuming an  
approximate buoyancy velocity of $\sim$0.6~\cs, where
\cs~is the sound speed \citep{chu01}, we
estimate a buoyancy speed of 270 \kms~in the 
0.77~keV medium \citep{iso06}. The X-ray cavities
are located 320$\arcsec$ from the nucleus and \citet{wad61} 
measured the separation of the radio lobes to be 33$\arcmin$. If we assume the
lobe expansion is in the plane of the sky, for NGC 1316's distance of 22.7~Mpc,
the X-ray cavities and the radio lobes are at 35~kpc and 108~kpc 
from the nucleus respectively. These distances correspond to buoyancy
rise times of 0.1~Gyr for the X-ray cavities and 0.4~Gyr 
for the radio lobes. The age of 0.4~Gyr appears large for radio lobes 
still emitting at 1.4~GHz, but matches the estimate of the
synchrotron age calculated by \citet{eke83}.
Further, if the lobe is continuously or intermittently connected to the nuclear
power supply, fresh injection of electrons or re-energization of
existing electron populations could, in principle, supply
sufficient high energy electrons so that the lobe is visible
at GHz frequencies even after several hundred million years. 
Finally, \citet{iyo98} estimated that the nucleus of NGC 1316
was active $\sim$0.1~Gyr ago, which agrees with our estimate of the
age of the X-ray cavities.

\citet{mac98} estimated a single merger with 
a low-mass gas-rich galaxy $\sim$0.5~Gyr ago could be responsible
for the optical tidal tail morphology. Such a merger could have
provided the material to power the AGN outbursts.
A gas-rich merger galaxy would contribute blue young stars,
which require about a Gyr to become red and dead. The B-V
color of NGC 1316 of 0.87 \citep{dev91} is slightly bluer 
than the typical B-V color of 0.91 for elliptical galaxies \citep{tri00}, 
which suggests that NGC 1316 may still contain a small population of 
early type stars. NGC 1316's color therefore supports a merger 
within the last Gyr. We also can set a lower limit on the merger
age by estimating the free-fall time of the northwestern dust component. 
We revised the \citet{arn98} total mass within 45$\arcsec$ of 
the nucleus to be $6.6 \times 10^{10}$~\msun~and used this along 
with the excess velocity of 70\kms~for the northwestern clump of molecular gas 
\citep{hor01} to estimate a free-fall time for the northwestern
dust and molecular gas feature. The estimated free-fall time 
of 22~Myr is a lower limit on the age of the merger, since 
the clump likely also has a tangential velocity component and was
likely deposited at greater radii by the merger.

\subsection{Outburst Energies}

\citet{chu02} described the energy deposition 
required to inflate a bubble adiabatically, and thereby create a
cavity, as the enthalpy of that bubble, which for
relativistic gas is $4PV$. To estimate the
energy of the outburst responsible for the X-ray cavities, we 
used the more clearly defined western X-ray cavity, 
whose shape we approximate as a sphere of 230$\arcsec$ 
(25~kpc) centered 320$\arcsec$ (34.8~kpc) 
from the nucleus. To measure the pressure, we
assume an isothermal gas at 0.77 keV \citep{iso06} and
solar abundance and derived the density from the 
surface brightness. We model the density as a 
$\beta$ model \citep{cav76}, whose parameters we derive by
fitting the surface brightness profile of the
exposure-corrected \xmm~image in a region not containing
the cavities. This method provides a 
lower limit on the total outburst energy, since it
estimates the kinetic energy released in the outburst.
Assuming solar  abundance, we estimate the kinetic outburst energy
is $10^{58}$~ergs, for equal-sized bubbles created in
the plane of the sky on each side of the nucleus. If the abundance is 
half solar, then the gas density and outburst energy 
both increase by $\sim$40~\%. Based on the energy  
needed to create the X-ray cavities and adopting a 
mass-energy conversion efficiency of $\epsilon = 10\%$, we
estimate the mass of material that would have been 
accreted onto the SMBH to be:
\begin{equation}
\Delta M_{BH} = \frac{(1-\epsilon)}{\epsilon} \frac{E}{c^{2}} = 5 \times 10^{4} M_{\odot}   
\end{equation}
where $E$ is the total energy output. 

Deep X-ray observations are not available for a similar analysis 
in the regions defined by the radio lobes, and we expect complications due
to inverse Compton X-ray emission coincident with the potential cavities
\citep{fei95}. However, if we extrapolate the gas density model 
to the radio lobes assuming they lie in the plane of the sky and use 
the 20~cm observation to determine the location and size of the lobes, 
we can estimate the energy required to evacuate cavities the present 
size of the lobes and thereby estimate the energy of the outburst 
required to create them. We approximated the lobes as 24$\arcmin$ (78.3~kpc) 
diameter spheres centered at 14\farcm3 (93.3~kpc) (west) and 
15\farcm6 (101.4~kpc) (east) from the nucleus. We extrapolated 
the gas density model derived from the \xmm~surface
brightness to the radii of the radio lobes and combined the derived pressure
there with the expected cavity volumes created through 
adiabatic bubble expansion to estimate the required energy of 
$\sim5\times10^{58}$~ergs. The mass accreted
onto the SMBH to produce this energy would be $\sim2\times10^{5}$~\msun.

\subsection{Comparison of the Cen A and Fornax A Jets}
Centaurus A (NGC 5128) and NGC 1316 (Fornax A) are both nearby
elliptical galaxies, which have recently undergone a merger
event that has produced strong nuclear activity. They each 
host a 10$^{8}$~M$_{\odot}$ black hole that are low luminosity AGNs and 
have dust lanes roughly perpendicular to their radio lobes
\citep{mar06, now08}. However, they 
differ significantly with regards to the observational
characteristics of their jets. At a distance of 3.7 Mpc
\citep{fer07}, Cen A has a 1.5~kpc radio and X-ray emitting 
jet that extends from the nucleus to the northeast radio lobe 
\citep{fei81, kra02, har03}. In contrast, NGC 1316's $\sim$3~kpc radio
jet does not extend to the large radio lobes and coincides 
not with an X-ray jet, but with a soft X-ray cavity. This phenomenological 
contrast suggests that, while the jet of Cen A is dissipative
\citep{nul10}, NGC 1316's jet
is not or that the NGC 1316 jet has shut off on larger scales. 

\section{Conclusions}
We detected considerably more dust emission for NGC 1316 than expected
in an early type galaxy with its K-band luminosity
and observed evidence of recent AGN outbursts in
the form of X-ray cavities and radio lobes.
We presented \spitzer~images of the infrared dust 
emission, including the first image of dust emission at 5.8\um. 
We determined that 
the dust has a clumpy morphology, mostly confined to two regions, one 
28\farcs8 (3.1~kpc) southeast of the nucleus and a 
43\farcs9 (4.8~kpc) protrusion ending in an arc northwest 
of the nucleus. Molecular emission is detected from these regions. 
The resolved radio jet is not detected by \spitzer, and it 
does not coincide with regions of dust emission. Since the dust must be
almost entirely external in origin, we use the dust mass to constrain
 the type and mass of the merger galaxy.  We calculated a present
dust mass of $2.4\pm 0.9\times 10^{7}$~\msun~based on the integrated MIPS
fluxes, which agrees with the dust mass of 
$3.2 \times 10^{7}$~\msun~predicted by the model of \citet{dra07}.
We estimate the merger galaxy was a late type galaxy with 
a stellar mass of $1-6\times10^{10}$~\msun~and a gas mass 
of $2-4\times10^{9}$~\msun, some of which was likely ionized or
used to form stars in the merger event.

We constrained the age and energy of the merger and outburst events based
on the X-ray and radio emission. The \xmm~image shows a pair of X-ray cavities
to the west and southeast of the nucleus, likely created by the expansion
of radio plasma, which are closer to the nucleus than the 1.4~GHz radio
lobes. The relative locations of these cavities and the radio lobes
suggests that either the 1.4~GHz radio emission does not show the full extent of
the radio emission from the outburst or that there have been at 
least two AGN outbursts. We calculate 
buoyant rise times for the X-ray cavities of 0.1~Gyr and for the radio lobes 
of 0.4~Gyr, assuming expansion in the plane of the sky at 0.6~\cs, 
which agrees with the synchrotron age estimated by \citet{eke83}. 
 Since the age of the radio lobes is close to the 
0.5~Gyr age estimated by \citet{mac98} for the merger, the outburst
was likely triggered by the accretion of material onto the SMBH 
from this merger. Finally, we constrained the kinetic energy of the
outbursts based on the energy required to 
create the \xmm~cavities and the radio lobes bubbles.
We estimate the outburst that created the X-ray cavities had a kinetic
energy of $10^{58}$~ergs and that the creation
of the radio lobes required $\sim5$ times more power.

\acknowledgements

We are grateful to C. Horellou and J.~H. Black for providing us with CO(2-1)
intensities and for their comments regarding the molecular gas kinematics, 
to Zhiyuan Li for his useful discussion on calculating 
masses, to Dharam Lal for his assistance in creating the
4.89~GHz map, and to Ramesh Narayan for his comments. We 
thank the anonymous referee for the many comments that
improved this work. This work was based on archival data obtained from the 
Spitzer Science Archive, the Chandra Data Archive, and the XMM-Newton
Science Data Archive. Archived images were also obtained from the
Hubble Legacy Archive, the NASA/IPAC Extragalactic Database, and the
National Radio Astronomy Observatory Archive. We thank Z. Levay of the
Space Telescope Science Institute for his assistance in obtaining the
Hubble ACS image. This work was supported by the Smithsonian
Institution, the Chandra X-ray Center, and NASA contract NNX07AQ18G.

{\it Facilities:} \facility{Spitzer}, \facility{CXO}, \facility{XMM}, \facility{VLA}, \facility{HST}

\end{document}